\documentclass[pdflatex,sn-mathphys,Numbered, smallextended]{sn-jnl}

\usepackage{graphicx}
\usepackage{amsmath,amssymb,amsfonts}
\usepackage{amsthm}
\usepackage{tikz}
\usetikzlibrary{patterns, fit, positioning, arrows.meta}
\usepackage{pgfplots}
\pgfplotsset{compat=1.16}
\usepackage{booktabs}
\usepackage{manyfoot}
\usepackage{hyperref}
\usepackage{xspace}
\usepackage{microtype}
\usepackage{tcolorbox}
\usepackage{algorithm}
\usepackage{algpseudocode}
\usepackage{url}
\usepackage{tabularx}

\newcommand{\codename}{\textsc{Aegis}\xspace}

\newtheorem{property}{Property}[section]

\newtheorem{invariant}{System Invariant}[section]

\begin{document}

\title[Evaluating Architectural Limits of Cloud TPUs]{Architectural Limits of Cloud TPUs in Finite-Field Cryptography}

\author*[1]{\fnm{Hung} \sur{Dang}}\email{hung.dk@vlu.edu.vn}
\author[2]{\fnm{Xuan Phu} \sur{Dang}}\email{26210059@ms.uit.edu.vn}
\author[3]{\fnm{Tue} \sur{Nguyen}}\email{tuent@appota.com}

\affil*[1]{\orgdiv{Van Lang School of Technology}, \orgname{Van Lang University}, \orgaddress{\country{Vietnam}}}
\affil[2]{\orgdiv{University of Information Technology}, \orgname{Vietnam National University}}
\affil[3]{\orgname{Appota}, \orgaddress{\country{Vietnam}}}

\abstract{We empirically characterise the cost-efficiency deficit between cloud Tensor Processing Units and GPUs for finite-field cryptography. Against A100 GPU baselines (cuZK), we measure a $[5{,}558\times, 6{,}908\times]$ deficit across v5p and v4 architectures under an FP32-mantissa staging discipline, and a $\sim$$4{,}693\times$ deficit using v5p's native \texttt{int32} accumulator. We analytically project this deficit into a fundamental arithmetic penalty (lacking wide-integer ALUs) and a spatial penalty. We demonstrate that evaluating concurrent multi-tenant deployments, where strict separation forces eager Montgomery reduction, yields a projected $5.19\times$ spatial collapse; relaxing this constraint theoretically recovers these spatial cycles, yet the underlying arithmetic penalty remains. To facilitate this characterisation, we deploy \codename as a measurement vehicle. By mapping low-degree polynomials onto matrix-form Number Theoretic Transforms, the scheduler stacks heterogeneous polynomials into dense 2D matrices, achieving $\sim$$100\%$ K-dimension column occupancy on uniform workloads ($>$$92\%$ on mixed-degree traces). However, despite optimal K-dimension packing, severe M-dimension under-utilisation (e.g., $6.25\%$ on v4) combined with overwhelming VPU-bound Montgomery reduction stalls mathematically starve the systolic arrays. A post-hoc HLO validator ensures these measurements remain structurally isolated against the XLA fusion engine. Our findings empirically demonstrate the structural inadequacy of AI-optimised systolic arrays for exact, high-throughput field arithmetic.}

\keywords{Tensor Processing Units, Multi-Tenant Scheduling, Number Theoretic Transform, Finite-Field Cryptography}

\maketitle

\section{Introduction}
\label{sec:introduction}

The Number Theoretic Transform (NTT), the asymptotic bottleneck of zero-knowledge proofs and post-quantum signatures, admits a dense matrix-form expression mapping naturally onto AI-optimised systolic arrays. This positions cloud Tensor Processing Units (TPUs)~\cite{TPU-V4, MORPH} as apparent successors to GPU-resident NTT accelerators~\cite{CUZK, GZKP}: the $128 \times 128$ Matrix Multiplication Unit (MXU) natively accelerates this dense formulation, bypassing the sparse radix-2 butterfly network of the asymptotically optimal $\mathcal{O}(d \log d)$ NTT. We show this apparent fit does not survive contact with real cryptographic arithmetic. Across three TPU generations (v4, v5e, v5p), AI-optimised systolic arrays exhibit a $[5{,}558\times, 6{,}908\times]$ cost-efficiency deficit against A100 GPU baselines for BN254 multiplication, remaining $\sim$$4{,}693\times$ behind even under v5p's native \texttt{int32} accumulator path.

Two structural tensions explain the gap: arithmetic mismatch and tenant separation. TPUs natively support \texttt{bfloat16} and \texttt{int8} formats, whereas cryptographic workloads demand exact arithmetic over disparate prime fields. Because the prime modulus exceeds the $2^8=256$ exact-integer ceiling of the \texttt{bfloat16} pipeline, post-quantum signatures like CRYSTALS-Dilithium ($Q = 8{,}380{,}417$, 23-bit precision) force a 3-limb \texttt{int8} decomposition. Legacy BN254 zero-knowledge proofs demand exact 254-bit arithmetic: the Explicit Residue Number System (ERNS) pipeline (Section~\ref{sec:implementation}) maps each field operation to nine 32-bit residues, further decomposed into 8-bit limbs. This yields 144 pointwise limb cross-products plus $>$$2{,}100$ base-extension multiplications for Montgomery reduction. Against this expansion, the systolic accumulator's per-pass capacity binds the schedule long before asymptotic matmul throughput becomes relevant.

Systolic accumulator precision strictly bounds the maximum single-pass polynomial degree before VPU intervention. Section~\ref{sec:correctness_bounds} formally derives that multi-limb arithmetic rapidly exhausts these bounds, precluding single-pass evaluation of Layer-2 ZK-rollup polynomials ($d \ge 2^{20}$) and forcing compiler-managed VPU re-injection. This structurally relegates TPUs to processing discrete, low-degree primitives. We establish the BN254 base hardware tile limit at $d \le 128$ (Dilithium at $d \le 171$), define the evaluation "op" at $d=256$, and scale multi-tenant testing to $d=8{,}192$. Multi-tenant sequencers process heterogeneous polynomial streams. Sequential dispatch chronically under-utilises the array's geometry, while naive batching risks cross-tenant memory collisions across shared High Bandwidth Memory (HBM), VPU registers, and accumulators.

We propose \codename, a two-tier scheduling architecture: Tier 1 implements a Rectangular Scheduler stacking heterogeneous tenant polynomials into dense $N_c \times \hat{d}_{\max}$ matrices (eliminating block-diagonal structural-zero waste), and Tier 2 implements a Slice-Level Co-Scheduler managing across-TensorCore dispatch. A cross-cutting Compiler-Enforced Workload Separation subsystem statically partitions HBM and forbids cross-workload HLO fusion (Section~\ref{sec:system_design}).

Our primary contribution establishes that AI-optimised systolic arrays are architecturally mismatched for exact, high-throughput finite-field cryptography. \codename serves as the measurement vehicle enabling this empirical characterisation, delivering two principal contributions:
\begin{enumerate}
    \item We formally and empirically delineate the architectural envelope of AI accelerators for cryptography. Building on Section~\ref{sec:correctness_bounds}, we report a controlled-baseline cost-efficiency deficit of $[5{,}558\times, 6{,}908\times]$ across v5p and v4 under a uniform FP32-mantissa staging discipline, and a sensitivity-run deficit of $\sim$$4{,}693\times$ when v5p's native \texttt{int32} accumulator path is enabled. Section~\ref{sec:evaluation} analytically projects this massive deficit into a fundamental arithmetic penalty (lacking wide-integer ALUs) and a multi-tenancy spatial penalty. This spatial penalty projection assumes our separation methodology: preventing cross-tenant memory aliasing among concurrent batch workloads necessitates strict row-separation, forcing eager Montgomery reduction. While relaxing this discipline in trusted, single-tenant environments theoretically recovers the spatial loss, the underlying projected arithmetic deficit remains.
    \item We expose the necessity of compiler-enforced mixed-workload validation in heterogeneous edge environments. Concurrently dispatching lightweight Dilithium arrays alongside heavy BN254 batches safely parallelises processing, but the XLA compiler aggressively optimises across workload boundaries. We develop a post-hoc HLO validator that statically asserts strict structural invariants on the lowered HLO, providing a deterministic correctness guarantee against the class of cross-tensor fusions that XLA's instruction fusion pass would otherwise permit.
\end{enumerate}

\section{Background}
\label{sec:background}

\textbf{Cryptographic Acceleration and TPUs.}
Zero-Knowledge Proofs (ZKPs) enable computational verification without witness disclosure. Modern systems (e.g., pairing-based SNARKs~\cite{GROTH16, PLONK}, transparent STARKs~\cite{STARK}) rely on polynomial commitments over large prime fields like BN254~\cite{BN254} or BLS12-381. Generating these proofs requires evaluating polynomials across large domains via the Number Theoretic Transform (NTT). The NTT's $\mathcal{O}(d \log d)$ radix-2 butterfly network maps to dense matrix multiplications over $\mathbb{F}_p$, rapidly bottlenecking CPUs and forcing migration to parallel hardware.
Google's Tensor Processing Units (TPUs) accelerate dense tensor operations via a core Matrix Multiplication Unit (MXU). The TPU v4 MXU, a $128 \times 128$ systolic array, is backed by High Bandwidth Memory (HBM). A distinct Vector Processing Unit (VPU) handles modular reductions essential for finite-field arithmetic. Crucially, TPUs lack hardware-enforced context boundaries or confidential computing modes~\cite{SEV-SNP, H100CC}; the XLA compiler manages memory statically without physical sub-graph isolation.

\textbf{Precision and Multi-Tenancy Gaps.}
Cryptographic protocols demand exact arithmetic over disparate prime fields, clashing with the TPU's native \texttt{bfloat16} and \texttt{int8} pipelines. Legacy SNARKs over BN254 require 254-bit arithmetic, forcing a heavy 4-limb \texttt{int8} decomposition per 32-bit residue. Conversely, post-quantum signatures like CRYSTALS-Dilithium operate over a 23-bit prime ($Q = 8{,}380{,}417$), requiring a 3-limb decomposition. As sequencers process continuous proof streams from disparate edge applications, a structural gap emerges: while AI natively saturates the MXU, individual client polynomials exhibit degrees $d_i \ll 128$. Sequential execution chronically under-utilises the array, while concurrent execution without structural segregation induces functional collisions across shared HBM, VPU registers, and systolic accumulators.
\section{Problem Formulation and System Constraints}
\label{sec:problem}

\subsection{System Model}
We model a multi-tenant cloud environment where decentralised tenants submit low-degree cryptographic polynomials for evaluation. A sequencer batches these requests and dispatches dense matrix-form Number Theoretic Transform (NTT) evaluations to a shared Tensor Processing Unit (TPU). 
Let $N_c$ denote the per-TensorCore batch depth (the M-dimension fill), and $N_s$ denote the slice-wide concurrent tenants ($N_s = N_c \times \text{number of TensorCores}$). Tenant $i \in \{1, \dots, N_c\}$ co-scheduled on a given TensorCore provides a public polynomial of unpadded degree $d_i$ over a prime field $\mathbb{F}_p$. The XLA-compatible padded degree is denoted $\hat{d}_i = \lceil d_i / d_{\max} \rceil \cdot d_{\max}$, where $d_{\max}$ is the native hardware staging capacity. 
To prevent wasting spatial array slots on structural zeros when padding $N_c$ polynomials into a monolithic block-diagonal operand, the sequencer maps constituent polynomial evaluations onto a stacked $N_c \times \hat{d}_{\max}$ matrix operand, $\mathbf{A}_{\text{stack}}$, for parallel systolic array execution against a shared twiddle matrix $\mathbf{W}$.

\subsection{Correctness Model and Constraints}
\label{sec:problem:threat_model}
We adopt a correctness model where the sequencer and XLA compiler execute the software stack as specified but may admit functional-correctness regressions from overly aggressive HLO fusion. This is a software-fault model, not a security adversary model: confidentiality, side-channel resistance, and Byzantine cloud-provider attacks are out of scope. However, we assume a concurrent multi-tenant execution environment (e.g., decentralised prover networks) where preserving functional correctness is paramount: one tenant's bug, pathological input, or XLA-induced compiler aliasing must strictly not corrupt another tenant's mathematical output.

Tenants submit raw polynomial coefficients via a restricted REST/RPC API; they do not submit arbitrary intermediate representations or HLO graphs. The sequencer retains exclusive control over the XLA graph generation, structurally guaranteeing isolated tenant computations. 
The primary systems challenge involves bin-packing divergent cryptographic requests onto a rigid systolic architecture while preventing cross-tenant data aliasing. \codename must satisfy three multi-tenancy constraints:
\begin{enumerate}
    \item \emph{Data Correctness}: The batched output returned to a tenant must remain strictly isomorphic to an isolated evaluation. Dense spatial stacking against a shared twiddle matrix must preclude cross-tenant memory pollution intrinsically.
    \item \emph{Type Homogeneity}: Sequencers aggregate substantial variance in polynomial degrees and precisions. Because physical TPU instructions demand strict type homogeneity, the compiler must structurally prevent cross-domain operation fusion to avert undefined hardware behaviour.
    \item \emph{State Pollution}: Shared execution units retain state, and TPUs lack hardware-level MXU context-switch support. The architecture must explicitly manage residual tenant data to prevent runtime artifacts from polluting co-tenant mathematical contexts.
\end{enumerate}

\section{Architecture: Two-Tier Scheduling and Workload Separation}
\label{sec:system_design}

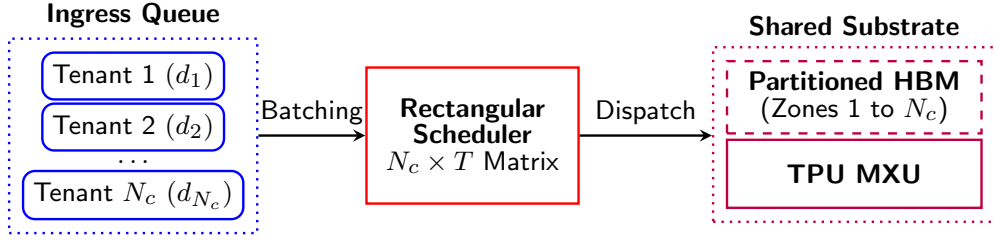
\begin{figure}[t]
    \centering
    \resizebox{\linewidth}{!}{%
    \begin{tikzpicture}[
        node distance=0.6cm,
        font=\sffamily\scriptsize,
        tenant/.style={draw=blue, thick, rounded corners, minimum width=1.5cm, minimum height=0.3cm, align=center},
        weaver/.style={draw=red, thick, pattern=north west lines, pattern color=red!30, minimum width=2.5cm, minimum height=1.6cm, align=center},
        mxu/.style={draw=purple, thick, minimum width=3cm, minimum height=0.8cm, align=center},
        hbm/.style={draw=purple, thick, dashed, minimum width=3cm, minimum height=0.6cm, align=center},
        arrow/.style={->, >=stealth, thick}
    ]

    % Tenants
    \node[tenant] (t1) at (0, 0.6) {Tenant 1 $(d_1)$};
    \node[tenant] (t2) at (0, 0) {Tenant 2 $(d_2)$};
    \node (dots) at (0, -0.4) {$\hdots$};
    \node[tenant] (tn) at (0, -0.8) {Tenant $N_c$ $(d_{N_c})$};
    \node[draw=blue, thick, dotted, fit=(t1) (tn), inner sep=0.15cm, label=above:{\textbf{Ingress Queue}}] (ingress) {};

    % Weaver
    \node[weaver, fill=none] (weaver) at (4.0, -0.1) {};
    \node[fill=white, text=black, inner sep=1pt, align=center] at (4.0, -0.1) {\textbf{Rectangular}\\\textbf{Scheduler}\\$N_c \times T$ Matrix};

    \node[hbm, fill=none] (hbm) at (8.5, 0.35) {\textbf{Partitioned HBM}\\(Zones 1 to $N_c$)};
    \node[mxu, fill=none] (mxu) at (8.5, -0.55) {\textbf{TPU MXU}};
    \node[draw=purple, thick, dotted, fit=(mxu) (hbm), inner sep=0.15cm, label=above:{\textbf{Shared Substrate}}] (hardware) {};

    \draw[arrow] (ingress.east |- weaver.west) -- node[above] {Batching} (weaver.west);
    \draw[arrow] (weaver.east) -- node[above] {Dispatch} (hardware.west |- weaver.east);

    \end{tikzpicture}%
    }

    \caption{\codename two-tier scheduling architecture.}

    \label{fig:architecture}
\end{figure}

Figure~\ref{fig:architecture} shows the two-tier scheduling architecture of \codename (Tier 1 Rectangular Scheduling and Tier 2 Slice-Level Co-Scheduling). The ingress queue buffers arriving tenant polynomials. Rather than dispatching low-degree polynomials sequentially, which severely under-utilises the systolic array, or constructing large block-diagonal operands that inflate complexity, the sequencer accumulates a batch of $N_c$ requests per TensorCore.
Because polynomials within a degree bucket share the twiddle matrix $\mathbf{W}$, the Rectangular Scheduler dynamically groups polynomials by length, pads each tenant's $1 \times d_i$ vector to the maximum bucket dimension, and stacks them, creating a dense $N_c \times \hat{d}_{\max}$ 2D matrix.
These are dispatched using XLA's native dense matrix multiplication (\texttt{jax.lax.dot}), mapping the $N_c$ independent requests directly to the $M$-dimension (spatial batch dimension) of the systolic array. The orchestrator subsequently transmits this combined tensor to the TPU.

\vspace{-2mm}
\subsection{Two-Tier Scheduling: Rectangular and Co-Scheduling}
\label{sec:system_design:weaver}

Because physical TPU MXU instructions demand strict data-type homogeneity within any given operand, \codename segregates incoming polynomials into workload-homogeneous queues.
For each workload class $C$, the Rectangular Scheduler constructs the stacked operand $\mathbf{A}_{\text{stack}}^{(C)} \in \mathbb{F}^{N_c \times \hat{d}_{\max}}$ row by row. It initialises a zero matrix, pads each tenant polynomial $p_i$ to the bucket dimension $\hat{d}_{\max}$, writes it into row $i$, and attaches the workload-zone tag $\mathcal{Z}_C$. Executing the batched multiplication operates in $\mathcal{O}(N_c \hat{d}_{\max}^2)$ time, physically saturating the array without the spatial waste of structural zeros caused by block-diagonal stacking.

The Slice-Level Co-Scheduler maps these batched tensors onto the physically distributed computing cores of a TPU pod slice, enabling concurrent execution of diverse cryptographic primitives. While stock \texttt{pmap} targets uniform execution across physical devices, \codename schedules heterogeneous workloads (e.g., Dilithium alongside BN254) across all available TensorCores simultaneously by tagging distinct workload streams at the logical \texttt{batch\_matmul} dimension. Strict logical isolation is enforced, mapping discrete tenant batches into disjoint HBM regions without demanding physical partition of the TensorCores themselves.

This separation follows from Property~\ref{prop:isolation}: row-dimension semantics of 2D matrix multiplication prevent cross-product interference $p_i(x) \cdot p_j(y)$ for $i \neq j$; the orchestrator emits a \texttt{workload\_zone} annotation for each row index $i$.

\subsection{Workload Separation and Context Management}
\label{sec:system_design:isolation}

While the rectangular structure compartmentalises tenant arithmetic, shared hardware state necessitates active management to avert functional collisions. \codename orchestrates spatial zoning, temporal context switching, and workload-class segregation via the XLA compiler.

\textbf{Spatial and Workload Zoning via XLA Semantics}
The TPU stages operands in HBM prior to systolic dispatch. HBM zone partitioning is implemented strictly to guarantee functional correctness against compiler-induced aliasing rather than as an arithmetic-correctness requirement; the \texttt{batch\_matmul} row semantics alone provide mathematical separation (Property~\ref{prop:isolation}).
However, unconstrained allocation permits the XLA memory planner to interleave tenant tensors spatially, and aggressive cross-workload graph optimisation risks fusing disparate structural expansions. To establish a strict correctness mechanism against compiler-induced aliasing, \codename statically partitions the compiler graph. This zoning explicitly prevents the XLA optimiser from invalidly fusing heterogeneous operations, such as merging 3-limb Dilithium blocks with 4-limb BN254 blocks, which would violate the hardware's strict precision homogeneity requirements.

The ingress sequencer implements this separation within the High Level Optimiser (HLO) representation, wrapping each tenant's batched tensor subset in explicit \texttt{xla::CustomCall} operations. These nodes carry \texttt{mhlo.custom\_call} dialect annotations defining a tenant-distinct \texttt{memory\_space}, a temporal \texttt{workload\_zone}, and a structural \texttt{precision\_zone}.~Constrained by these metadata attributes, the optimisation barriers prevent XLA from fusing operations across workload boundaries, aliasing memory between tenants, or merging distinct precision classes. \codename executes a structural validation pass prior to dispatch (Section~\ref{sec:implementation:validator}) to enforce these constraints against potential upstream XLA regressions. \\

\textbf{Temporal Serialization and Hardware Limitations}
Current TPU architectures lack hardware context-switch support for saving and restoring isolated MXU and Vector Processing Unit (VPU) states. Hardware-enforced temporal separation remains impossible without silicon modifications. \codename relies on spatial zoning and runtime serialisation. 
The XLA compiler annotations provide strict functional correctness against accidental aliasing, fusion artifacts, and cross-tenant state pollution. \codename makes no claims regarding execution timing. 
The HBM partitioning and workload-zoning mechanisms target \emph{functional} (cross-tenant arithmetic) correctness: they exist strictly to guarantee that compiler optimisations cannot interleave heterogeneous mathematics. Analysing or mitigating potential hardware-level resource contention (e.g., HBM bank saturation) requires infrastructure controls that remain out of scope for this architectural characterisation study (Section~\ref{sec:problem:threat_model}).

\section{Arithmetic Exactness and Algorithmic Crossover}
\label{sec:correctness_bounds}

We characterise \codename's functional isolation guarantees and theoretical utilisation improvements. We establish arithmetic exactness (Property~\ref{prop:isolation}) and compiler-mediated separation (Invariant~\ref{inv:compiler}), concluding with a scheduling efficiency model.

\subsection{Arithmetic Block Isolation and Exactness}
\label{sec:sec_analysis:block}

To establish the correctness of \codename, we must decouple the structural independence of the Rectangular Scheduler from the finite-field exactness of the TPU's physical accumulators. 

\begin{property}[Accumulator Exactness Bound]
\label{prop:isolation}
Under the parameters of \codename, the partial-sum accumulator preserves the exact integer multi-limb expansion of $p_i(x) \cdot \mathbf{W}$ within the mantissa window, so reducing modulo $q_i$ post-accumulation reproduces the field-correct result, provided the unpadded polynomial degree $d_i$ satisfies $d_i \le d_{\max}$, where $d_{\max}$ is bounded by the physical accumulator width.
\end{property}

\noindent\textbf{Justification.} By the formal definition of 2D matrix multiplication, the output row $\mathbf{c}_i$ is strictly $\mathbf{v}_i \mathbf{W}$ over the real numbers; computing row $i$ reads exclusively from row $i$ of the left operand, maintaining geometric independence inherently. \codename dispatches the \texttt{u8} $\times$ \texttt{s8} kernel via XLA's \texttt{DotGeneral} with \texttt{preferred\_element\_type=int32}, the lowering path documented in the AQT library\footnote{AQT: \url{https://github.com/google/aqt}}.
On TPU v4, this path specifically materialises partial sums through the MXU's FP32 path; the effective bound on exact integer representability is therefore the IEEE~754 single-precision mantissa, $2^{24} = 16{,}777{,}216$, not the nominal $2^{31}-1$ \texttt{s32} ceiling. We thus derive $d_{\max}$ against the mantissa-bounded staging window, and rely on tile-level VPU re-injection to extend evaluation beyond a single staging pass. Table~\ref{tab:accbench} provides empirical confirmation of this AQT-specific lowering behaviour.
Each \texttt{u8} $\times$ \texttt{s8} limb cross-product is bounded in magnitude by a maximum pixel-product of $255 \times 128 = 32{,}640$. The accumulator bound depends strictly on how these cross-products map to hardware. Rather than dispatching $C^2$ independent \texttt{DotGeneral} kernels (which would severely under-utilise the matrix unit), \codename geometrically interleaves the limbs of both operands into a single, fused $M \times K$ dense matrix multiplication. This spatial packing forces the MXU K-dimension to accumulate the multi-limb convolution directly. Consequently, the densest convolution diagonal forces the hardware accumulator to absorb exactly $C$ overlapping cross-products concurrently per output coefficient. A single output coefficient thus accumulates a maximum value of $C \times 32{,}640$ per polynomial degree, against which the mantissa ceiling $2^{24}$ admits a maximum unpadded degree of $d \le \lfloor 2^{24} / (C \times 32{,}640) \rfloor$. For Dilithium (3-limb $\times$ 3-limb), the analogous bound yields $d_{\max}^{\text{Dil}} = \lfloor 2^{24} / (3 \times 32{,}640) \rfloor = 171$; the operational $d = 256$ case therefore proceeds through two staging passes ($171 + 85$), with VPU reduction at the mantissa boundary.

For BN254 ($C=\text{4-limb \texttt{int8}}$ per 32-bit ERNS residue segment), the 4-limb structure supplies $4 \times 8 = 32$ bits. Tracing the densest diagonal of the 4-limb $\times$ 4-limb expansion within the single fused \texttt{DotGeneral} kernel, exactly 4 cross-products of matching modular weight collide in the same hardware accumulator. The mantissa-bounded staging limit on TPU v4 is therefore:
\[ d_{\max}^{\text{BN}} = \lfloor 2^{24} / (4 \times 32{,}640) \rfloor = 128. \]
A single 32-bit residue of a BN254 multiplication therefore admits exactly $d=128$ within one mantissa-safe pass, precisely matching the $128 \times 128$ MXU tile width. Beyond a single tile, \codename relies on compiler-managed staging: after each tile the VPU folds the partial sum back to a reduced field representative before the next tile dispatches. The full BN254 field product spans 9 residues and a Montgomery reduction introducing $>$ $2{,}100$ additional base-extension cross-products per coefficient; both phases are processed iteratively via the Vector Processing Unit (VPU) and streamed back to the MXU. 
The absolute end-to-end degree ceiling of a BN254 polynomial is therefore dominated by VPU capacity and HBM bandwidth long before the per-pass $d_{\max}^{\text{BN}}=128$ matters in isolation. 
Layer-2 ZK-rollup degrees ($d \sim 10^6$) remain architecturally unviable without continuous off-chip streaming; edge polynomials ($d \le 8{,}192$) remain safe under the compiler-managed staging discipline. \hfill$\square$

\begin{table}[h]
\caption{Empirical precision limits for the \texttt{DotGeneral} accumulator path under \texttt{u8} $\times$ \texttt{s8} lowering. TPU v4 rounds beyond the FP32 mantissa ($2^{24}$), whereas v5e/v5p natively support INT32 accumulation.}
\label{tab:accbench}
\centering
\begin{tabular}{lccccccc}
\toprule
\textbf{Target Partial Sum $S$} & $2^{23}$ & $2^{24}-1$ & $2^{24}$ & $2^{24}+1$ & $2^{25}-1$ & $2^{28}$ & $2^{30}$ \\
\midrule
TPU v4-8 Accumulator Exact & \textcolor{green!70!black}{\checkmark} & \textcolor{green!70!black}{\checkmark} & \textcolor{green!70!black}{\checkmark} & \textcolor{red!70!black}{$\times$} & \textcolor{red!70!black}{$\times$} & \textcolor{red!70!black}{$\times$} & \textcolor{red!70!black}{$\times$} \\
TPU v5e-8 Accumulator Exact & \textcolor{green!70!black}{\checkmark} & \textcolor{green!70!black}{\checkmark} & \textcolor{green!70!black}{\checkmark} & \textcolor{green!70!black}{\checkmark} & \textcolor{green!70!black}{\checkmark} & \textcolor{green!70!black}{\checkmark} & \textcolor{green!70!black}{\checkmark} \\
TPU v5p-8 Accumulator Exact & \textcolor{green!70!black}{\checkmark} & \textcolor{green!70!black}{\checkmark} & \textcolor{green!70!black}{\checkmark} & \textcolor{green!70!black}{\checkmark} & \textcolor{green!70!black}{\checkmark} & \textcolor{green!70!black}{\checkmark} & \textcolor{green!70!black}{\checkmark} \\
\bottomrule
\end{tabular}
\end{table}

\textbf{TPU v5e/v5p Accumulator Bound.} TPU v5e and v5p provide a native \texttt{int8} multiplier path paired with a true \texttt{int32} accumulator, extending the per-pass capacity for the 4-limb BN254 pipeline to $d_{\max}^{\text{BN}} = 16{,}448$ under the relaxed $2^{31}-1$ ceiling. We maintain the FP32-mantissa-bounded staging discipline uniformly across all measured generations to ensure a controlled architectural baseline, additionally conducting a v5p sensitivity run on the native \texttt{int32} path (Section~\ref{sec:evaluation}). Because the physical $128 \times 128$ MXU geometry persists across generations, VPU-managed tile reduction remains mandatory under both regimes, deterministically migrating the bottleneck from accumulator width to vector unit and HBM bandwidth.

Property~\ref{prop:isolation} establishes arithmetic separation \emph{in the algebraic model} up to the staging-window limit. The bound rests on the specific accumulator semantics of the target MXU under the AQT \texttt{int32}-preferred lowering, and assumes the compiler completes one precision-safe staging pass before VPU re-injection. 
Translating this algebraic guarantee into a memory-level execution guarantee requires the XLA lowering to neither prematurely fuse Montgomery reduction into the pointwise phase nor coalesce staging passes across the precision boundary. 

As detailed in Section~\ref{sec:system_design}, the \codename JAX frontend wraps each tenant segment in an \texttt{xla::CustomCall} node carrying distinct \texttt{mhlo.memory\_space} and \texttt{precision\_zone} annotations, separated by explicit optimisation barriers. XLA's static shape inference structurally discharges four background conditions on the compiled module ($\mathcal{M}^\star$): it ensures disjoint addressing between tenants, prevents cross-block fusion of distinct memory zones, avoids alias liveness extension across zones, and enforces strict workload-zone separation (preventing fusion of mismatched precision classes). 

With these spatial and structural boundaries guaranteed by the compiler, the functional correctness of the multi-limb execution depends entirely on preserving the exact sequence of pointwise expansion and base-extension. We characterise this final premise as a strict operational invariant.

\begin{invariant}[Strict Reduction Ordering]
\label{inv:compiler}
VPU Montgomery reduction operations are never scheduled within the index range of an open pointwise summation loop.
\end{invariant}

Invariant~\ref{inv:compiler} describes an operational requirement of the XLA implementation. The \codename artifact (Section~\ref{sec:implementation}) includes a structural validator that empirically asserts this condition (alongside the four background compiler assumptions) on every compiled module prior to dispatch; any violation triggers a dispatch abort. Property~\ref{prop:isolation} and Invariant~\ref{inv:compiler} together ensure that algebraic independence translates into memory-level execution independence and that the temporal-barrier discipline survives compilation.

\subsection{Utilization Modeling and Algorithmic Crossover}
\label{sec:sec_analysis:utilisation}

To quantify \codename's architectural deficit relative to an optimal Number Theoretic Transform (NTT), we define the effective utilisation $U_{\text{eff}}$ as the product of the geometric K-dimension MXU occupancy ($S_{\text{mxu}}$) and the algorithmic efficiency penalty ($P_{\text{algo}}$). While batched $\mathcal{O}(Nd^2)$ dense matrix multiplication geometrically saturates the systolic array ($S_{\text{mxu}} > 92\%$) by eliminating block-diagonal structural zeros, it remains an asymptotically redundant $\mathcal{O}(d^2)$ operation compared to the $\mathcal{O}(d \log d)$ NTT. This algorithmic penalty scales as $P_{\text{algo}} \approx (\log_2 d)/d$. For $d=256$, $P_{\text{algo}} \approx 0.03$, yielding an effective algorithmic throughput of $U_{\text{eff}} \approx 2.8\%$. The matrix-form NTT decays as $\mathcal{O}(d^2)$ while the GPU's bandwidth-bound radix-2 NTT scales as $\mathcal{O}(d \log d)$. No crossover degree exists at $d \ge 256$; the deficit grows monotonically with $d$.

\section{JAX/XLA Realisation and HLO Validation}
\label{sec:implementation}

We implemented \codename on Google Cloud Platform (GCP), leveraging a custom JAX/XLA stack to enforce microarchitectural constraints and orchestrate high-throughput cryptographic arithmetic.

\subsection{Execution Environment and Software Stack}
We deploy on TPU v4-8, v5e-8, and v5p-8 slices, each exposing 8 logical TensorCores; the $128 \times 128$ MXU geometry is uniform across generations~\cite{TPU-SYS-ARCH}.

The software stack couples a pinned JAX commit with a pinned XLA HEAD commit.\footnote{JAX: \url{https://github.com/jax-ml/jax/commit/fae53cce85c9d4ba558bad7b6e64be46f0545cfd}; XLA: \url{https://github.com/openxla/xla/tree/9eb2e3662a439d9030cbce26dc2f89b359e023b5}. The initial Ahead-Of-Time (AOT) compilation of the statically shaped composite graph incurs a $\sim$140-second cold-start penalty, which amortises over subsequent invocations and is mitigable via warm compilation pools. This one-time AOT compilation cost is strictly excluded from all reported throughput and latency measurements.}
Pinning the compiler ensures deterministic memory-planning and buffer-assignment heuristics across evaluation runs, which is critical for verifying HBM zone integrity. The Rectangular Scheduler, implemented via JAX transformations, constructs the stacked dense tensor $\mathbf{A}_{\text{stack}}$. Each dispatch evaluates a naive matrix-form Number Theoretic Transform (NTT) in $\mathcal{O}(d^2)$ time. Although asymptotically inferior to the $\mathcal{O}(d \log d)$ Cooley-Tukey NTT, evaluating the matrix-form NTT structurally maximises systolic array utilisation, trading algorithmic complexity for hardware parallelism. A suite of HLO-level annotations and post-hoc validator passes at the MLIR~\cite{MLIR} dialect layer implements the Compiler-Enforced Workload Separation. Explicit \texttt{xla::CustomCall} operations carrying custom \texttt{mhlo} dialect annotations enforce spatial zoning, constraining the memory planner to allocate disjoint physical HBM regions for each tenant partition. 
Custom workload metadata (\texttt{precision\_zone}, \texttt{workload\_zone}) is carried as opaque integer payloads on \texttt{xla::CustomCall} operands; the Python validator extracts these payloads from the lowered HLO post-buffer-assignment. The artifact includes a CPU-reference differential tester confirming arithmetic correctness of all batched evaluations.

\subsection{Cryptographic Arithmetic}
\codename instantiates polynomial evaluation over the BN254 elliptic curve prime field, adapting the Explicit Residue Number System (ERNS) pipeline from cuZK~\cite{CUZK} and GZKP~\cite{GZKP}. ERNS maps the 254-bit prime into a nine-residue moduli chain (eight base residues plus an auxiliary residue for overflow handling). This auxiliary residue is mandatory: whereas cuZK absorbs base-extension growth into native 128-bit GPU registers, \codename's accumulator strictly caps at the 24-bit FP32 mantissa. An 8-residue chain would overflow this bound during Montgomery reduction, requiring the 9th residue to constrain cross-products safely. 
Because the MXU accepts only 8-bit integer (\texttt{int8}) or 16-bit floating-point inputs, dispatching 32-bit integers forces a severe Vector Processing Unit (VPU) fallback penalty. 
\codename decomposes 32-bit residues into four 8-bit limbs, routing kernels through XLA's \texttt{DotGeneral} with \texttt{preferred\_element\_type=int32} (the AQT-documented \texttt{u8} $\times$ \texttt{s8} lowering). A single 32-bit point generates 16 partial cross-products. \codename stages each window at the FP32 mantissa bound (Property~\ref{prop:isolation}). Across the 9-residue chain, the BN254 point multiplication phase consumes $9 \times 16 = 144$ cross-products.
Full field multiplication requires a Montgomery reduction to enforce field bounds, introducing over $2{,}100$ base-extension cross-products per BN254 multiplication. Evaluating exact ERNS Montgomery reduction requires two dense base-extension matrix-vector multiplications across 8 primary residues, mapping to $2{,}048$ limb-level operations, plus 9th-residue scaling overhead that pushes the total strictly above $2{,}100$~\cite{CUZK}. The hardware artifact measures both the point multiplication phase (where geometric MXU packing is recovered) and end-to-end reduction.
The end-to-end BN254 throughput reported in Section~\ref{sec:evaluation} ($3{,}663$ ops/sec) is directly measured by running the complete pipeline (staging passes, in-GEMM re-injection, and Montgomery reduction) as one timed kernel. We treat the VPU-and-overhead penalty $\Pi$ as a derived diagnostic ratio, not an extrapolating multiplier.
Diagnostic penalty values ($\Pi$ and $\Pi_{\text{cf}}$) are derived directly in Section~\ref{sec:eval:utilisation} from isolated hardware measurements. Multi-tenant arrival penalty is independently measured by comparing the full pipeline at $N_s=1$ versus $N_s=8$, yielding a real $1.20\times$ overhead reflecting concurrent VPU stalls and HBM contention.

\subsection{HLO Structural Validation}
\label{sec:implementation:validator}

While the JAX frontend generates distinct memory space and precision zone annotations via \texttt{xla::CustomCall}, stock upstream XLA retains the authority to aggressively fuse, pad, and reallocate buffers during its optimisation pipeline. Because \codename operates on a standard, unmodified XLA backend, verifying that the compiler honors these separation boundaries natively requires an independent audit mechanism.

To guarantee functional correctness without deploying a custom XLA compiler fork, \codename implements a strict, post-hoc Structural HLO Validator in Python. This validator intercepts the fully compiled, lowered HLO module ($\mathcal{M}^\star$) just prior to hardware execution and statically asserts the invariants defined in Invariant~\ref{inv:compiler}.

The validator asserts Invariant~\ref{inv:compiler} (strict reduction ordering) alongside the four background conditions discharged by XLA's static shape inference---disjoint HBM addressing, cross-block fusion prevention, alias liveness containment, and workload-zone separation---as established in Section~\ref{sec:correctness_bounds}.

By executing this pass post-hoc over the stock XLA compiler output, \codename enforces structural correctness deterministically. Aggressive HLO fusion algorithms theoretically threaten these boundaries, as the stock XLA \texttt{instruction\_fusion} pass attempts a class of cross-tensor memory aliasing optimisations designed to maximize VPU utilisation. Without structural assertions, the compiler might coalesce the tail-end VPU Montgomery reduction instructions of one tenant with the initial \texttt{DotGeneral} pointwise expansion of a temporally adjacent tenant under mixed-precision schedules. Any assertion failure triggers a dispatch abort and emits the offending subgraph for triage, forcing a JIT recompilation with more restrictive \texttt{optimisation\_barrier} placement. Concretely, the JAX frontend emits a \texttt{jax.lax.optimisation\_barrier} after each tile of width $d_{\max}$, which prevents XLA's \texttt{instruction\_fusion} pass from coalescing adjacent passes and forces the compiler to schedule a VPU reduction op between them; the post-hoc validator asserts the presence of these barriers in the lowered HLO.
The validation step adds $\sim$$4.5$\,ms to the JIT compilation overhead per module, amortised across the JIT trace's lifetime and incurring zero per-dispatch runtime latency.

\section{Evaluation}
\label{sec:evaluation}

The evaluation quantifies the systemic trade-off between strict multi-tenant separation and batched computational throughput on matrix hardware. We explicitly define the unit-of-work ("op") to ensure valid cross-architecture comparisons.
For BN254 ($d=256$), one op is defined strictly as a single coefficient-wise full-field polynomial multiplication. On the TPU, this spans the entire end-to-end pipeline\footnote{All TPU measurements span Host~VM ingestion to HBM completion. \texttt{xprof} traces show PCIe ingress constitutes $<0.02\%$ of amortised wall-clock, dominated by the $>260\,\mu$s VPU compute bound, confirming the deficit is structural, not I/O-bound.}: the dual staging passes, in-GEMM matrix multiplication, and post-GEMM $>$$2{,}100$ base-extension Montgomery reduction. 
For the cuZK GPU baseline, one BN254 op refers precisely to the full-field coefficient-wise multiplication phase measured via cuZK's batched API. For Dilithium ($Q=8380417, d=256$), one op is defined as a single forward NTT. 
The ICICLE GPU proxy (32-bit scalar field) and the TPU measurement both report the throughput of this discrete forward NTT. Evaluating across three TPU generations (v4, v5e, and v5p), we performed over $250{,}000$ discrete matrix evaluations to address four research questions:
\begin{description}
\item[\textbf{RQ1}] What end-to-end matrix-form NTT throughput does \codename achieve for multi-limb workloads relative to state-of-the-art open-source GPU baselines (ICICLE (32-bit scalar field, Dilithium proxy) and cuZK (BN254))?
\item[\textbf{RQ2}] Where does the empirical architectural crossover point ($d_{\text{cross}}$) occur when scaling polynomial degrees against asymptotically optimal GPU NTTs?
\item[\textbf{RQ3}] What is the per-component overhead of the Compiler-Enforced Workload Separation?
\item[\textbf{RQ4}] How does the slice-level co-scheduler perform under heterogeneous, mixed-workload edge arrivals (e.g., Dilithium and BN254)?
\end{description}

All TPU experiments encompass a minimum of 10 trials per configuration. GPU baselines (Section~\ref{sec:eval:headline}) use a larger trial count of 30 because their per-trial cost is substantially lower. 
Reported point estimates carry 95\% bootstrap confidence intervals computed over $10^4$ resamples. The public evaluation scripts deterministically fix all randomisation seeds and record sample counts.

\subsection{Headline Throughput and Architectural Deficit}
\label{sec:eval:headline}
We benchmarked \codename against a sequential single-tenant baseline (where the TPU natively processes individual polynomials), a functionally equivalent time-sliced architecture, and cost-normalised A100 GPU references. To ensure rigorous benchmarking of the GPU side, we deployed the state-of-the-art open-source \texttt{icicle}\footnote{ICICLE: \url{https://github.com/ingonyama-zk/icicle/tree/625532a624e5aaa6e9d31a1c92587f1fcc30dc76}} and \texttt{cuZK}~\cite{CUZK} libraries on a Google Cloud \texttt{a2-highgpu-1g} instance. 
Because ICICLE does not ship a Dilithium NTT over $Q=8{,}380{,}417$, we use its 32-bit scalar-field NTT as a computational proxy for an optimistic upper bound on equivalent $\mathcal{O}(d \log d)$ bandwidth-bound throughput. For BN254 we map one ``op'' to \texttt{cuZK}'s batched evaluation API. 
GPU measurements reflect batched monolithic kernel execution inclusive of host--device PCIe overheads, with reported throughput as the median of 30 trials after a 5-run warm-up; 95\% bootstrap CIs are computed over $10^4$ resamples on per-run latencies. Even if software-level separation were imposed via CUDA streams, the A100's (40GB HBM2 variant) $\sim$1.555~TB/s HBM bandwidth~\cite{A100SPEC} executing the $\mathcal{O}(d \log d)$ NTT dominates the TPU. 
We therefore report the BN254 cost-efficiency deficit against the cuZK baseline ($7.2$M ops/sec).

We compiled cuZK with \texttt{CUDA 12.1} using the \texttt{nvcc -O3 -arch=sm\_80} flags to establish the primary BN254 baseline. To compile \texttt{icicle} for the Dilithium proxy, we used \texttt{cmake -S . -B build\_m31 -DFIELD=m31}.
The 32-bit-field Dilithium proxy benchmark used \texttt{NTTConfig\{batch\_size = 1024, are\_inputs\_on\_device = false\}}, ensuring host-to-device PCIe transfer is included in the measurement on the same basis as the cuZK BN254 benchmark.
The end-to-end measurement harness ($\approx$40 lines) is archived under \texttt{baselines/} in the artifact.

\begin{figure}[t]
    \centering
    \begin{minipage}[t]{0.468\textwidth}
        \vspace{0pt}
        \centering
        \begin{tikzpicture}
        \begin{axis}[
            width=0.85\linewidth, height=3.2cm, scale only axis,
            xmode=log,
            ymode=log,
            log basis x={2},
            log basis y={10},
            xtick={1, 2, 4, 8, 16, 32, 64, 128},
            xticklabels={1, 2, 4, 8, 16, 32, 64, 128},
            xlabel={Concurrent Tenants ($N_s$)},
            ylabel style={yshift=-0.2cm},
            ylabel={Throughput (ops/sec)},
            legend style={at={(0.5, 1.05)}, anchor=south, legend columns=-1, font=\footnotesize},
            grid=major,
            mark options={solid}
        ]
        
        \addplot[
            color=red, thick, mark=square, draw=red, fill=none,
            error bars/.cd, y dir=both, y explicit
        ] coordinates {
            (1, 3400) +- (0, 150)
            (2, 3400) +- (0, 150)
            (4, 3400) +- (0, 150)
            (8, 3400) +- (0, 150)
            (16, 3400) +- (0, 150)
            (32, 3400) +- (0, 150)
            (64, 3400) +- (0, 150)
            (128, 3400) +- (0, 150)
        };
        \addlegendentry{Sequential}

        \addplot[
            color=blue, thick, mark=triangle, draw=blue, dashed, fill=none,
            error bars/.cd, y dir=both, y explicit
        ] coordinates {
            (1, 3300) +- (0, 180)
            (2, 3300) +- (0, 180)
            (4, 3300) +- (0, 180)
            (8, 3300) +- (0, 180)
            (16, 3300) +- (0, 180)
            (32, 3300) +- (0, 180)
            (64, 3300) +- (0, 180)
            (128, 3300) +- (0, 180)
        };
        \addlegendentry{Zero-Padded}

        \addplot[
            color=teal, thick, mark=*, draw=teal, fill=none,
            error bars/.cd, y dir=both, y explicit
        ] coordinates {
            (1, 3400) +- (0, 150)
            (2, 6800) +- (0, 250)
            (4, 13600) +- (0, 450)
            (8, 27200) +- (0, 750)
            (16, 54400) +- (0, 1050)
            (32, 110000) +- (0, 1200)
            (64, 110000) +- (0, 1350)
            (128, 110000) +- (0, 1400)
        };
        \addlegendentry{\codename}
        \end{axis}
    \end{tikzpicture}
    \end{minipage}\hspace{2mm}
    \begin{minipage}[t]{0.468\textwidth}
        \vspace{0pt}
        \centering
        \begin{tikzpicture}
\begin{axis}[
    name=main,
    width=0.85\linewidth,
    height=3.2cm, scale only axis,
    xmode=log,
    ymode=log,
    log basis x={2},
    log basis y={10},
    xlabel={Polynomial Degree ($d$)},
    ylabel style={yshift=-0.2cm},
            ylabel={Throughput (ops/sec)},
    legend style={at={(0.5,1.05)}, anchor=south, legend columns=-1, font=\footnotesize},
    grid=both,
    grid style={line width=.1pt, draw=gray!30},
    major grid style={line width=.2pt,draw=gray!50},
    xtick={256, 1024, 4096, 16384},
    xticklabels={$2^8$, $2^{10}$, $2^{12}$, $2^{14}$},
    ymin=1, ymax=100000000
]

% GPU SPPARK Data
\addplot[
    color=blue,
    mark=square,
    line width=1pt,
    error bars/.cd, y dir=both, y explicit
] coordinates {
    (256, 18432600) +- (0, 600000)
    (1024, 3810200) +- (0, 140000)
    (4096, 824500) +- (0, 35000)
    (16384, 168200) +- (0, 7000)
};
\addlegendentry{SPPARK}

% cuZK Data (calibration point)
\addplot[
    color=orange,
    mark=*,
    mark size=3pt,
    only marks,
    error bars/.cd, y dir=both, y explicit
] coordinates {
    (256, 7200000) +- (0, 400000)
};
\addlegendentry{cuZK}

% TPU Aegis Data
\addplot[
    color=red,
    mark=triangle,
    line width=1pt,
    error bars/.cd, y dir=both, y explicit
] coordinates {
    (256, 3663) +- (0, 180)
    (1024, 245) +- (0, 12)
    (4096, 12) +- (0, 1)
    (16384, 0.8) +- (0, 0.04)
};
\addlegendentry{\codename}

\end{axis}
\end{tikzpicture}
    \end{minipage}
    
    \vspace{0.1cm}
    \begin{minipage}[t]{0.468\textwidth}
        \caption{Throughput scaling under concurrency ($d=256$ Dilithium).}
        \label{fig:throughput}
    \end{minipage}\hspace{2mm}
    \begin{minipage}[t]{0.468\textwidth}
        \caption{Empirical throughput scaling ($U_{\text{eff}}$) across degrees (BN254).}
        \label{fig:crossover}
    \end{minipage}
\end{figure}

Figure~\ref{fig:throughput} and Table~\ref{tab:cost_normalised} illustrate the absolute hardware performance. As shown in Figure~\ref{fig:throughput}, the throughput of the Rectangular Scheduler scales near-linearly at low concurrency ($N_s < 32$), reflecting the exact batch-dimension amortisation of the dominant VPU overhead, but effectively plateaus prior to reaching the full $128 \times 128$ tile limit because VPU reduction saturation bounds the pipeline before the geometric MXU capacity is entirely saturated.

For Dilithium workloads ($Q=8380417$), \codename requires a 3-limb \texttt{u8} $\times$ \texttt{s8} decomposition due to the $2^8=256$ exact-integer representability of the TPU's \texttt{bfloat16} format. \codename yields a measured end-to-end throughput of $110{,}435 \pm 1{,}200$ ops/sec at peak batch saturation on a TPU v4-8 pod slice. While the Rectangular Scheduler eliminates the spatial padding waste of block-diagonal scaling and dominates the TPU unbatched baseline (Table~\ref{tab:ablation}), it remains entirely uncompetitive. 
The A100 running cuDilithium~\cite{cuDilithium} achieves $1{,}409$K end-to-end Dilithium2 verifications/sec. As a Dilithium2 verification (ML-DSA-44, $k=l=4$) requires 13 NTT operations ($l=4$ forward NTTs for $\mathbf{z}$, $k=4$ forward NTTs for $\mathbf{t}_1$, $1$ forward NTT for $c$, and $4$ inverse NTTs to recover $\mathbf{w}'$)~\cite{FIPS204}, this yields an aggregate pipeline throughput of $\sim$$18.3\text{M}$ NTT ops/sec. We also evaluated the ICICLE \texttt{m31} 32-bit scalar-field NTT as an absolute optimistic upper bound for $\mathcal{O}(d \log d)$ memory bandwidth saturation, yielding $62.15\text{M}$ ops/sec. However, because $2^{31}-1$ is a Mersenne prime allowing near-free reduction, it overstates the throughput of generic primes like Dilithium's $Q = 8{,}380{,}417$ by $2\times$ to $4\times$ (as empirically observed in generic versus Mersenne-prime GPU microbenchmarks, e.g., \cite{CUZK}). We therefore rely on the cuDilithium projection for architectural comparisons. 
Since this pipeline execution time is shared with Keccak hashing, rejection sampling, and bit-packing, the GPU performs these $18.3\text{M}$ NTTs/sec while burdened by non-trivial cryptographic overhead. If the NTT operations consume a fraction $f \in [0.4, 0.8]$ of the GPU's verification wall-clock time, the true isolated GPU NTT throughput scales inversely to $18.3\text{M}/f$, yielding $22.8\text{M}$ to $45.7\text{M}$ ops/sec. The derived $18.3\text{M}$ figure therefore establishes a strict, conservative lower bound on the GPU's capabilities. Cost-normalised against this conservative projection, the A100 achieves $4.99$M ops/\$/hr versus the v5p-8 slice's $9{,}811$ ops/\$/hr (Table~\ref{tab:cost_normalised}), yielding a measured ${\sim}508\times$ cost-efficiency deficit for the Dilithium workload on the most recent v5p generation ($4.99\text{M} / 9{,}811$), with v4 yielding a wider ${\sim}582\times$ deficit. Because this baseline assumes $f=1$, the reported ${\sim}508\times$ deficit is a strict lower bound; realistic fractional usage would only widen the gap. While less pronounced than the BN254 deficit (${\sim}6{,}908\times$), reflecting the lighter 3-limb decomposition versus BN254's 16-cross-product inner expansion, both deficits reinforce the same structural conclusion: the systolic array's arithmetic mismatch with finite-field cryptography produces architectural under-competitiveness regardless of polynomial type.

Transitioning from the 3-limb Dilithium profile to the heavier 254-bit prime field, the BN254 workload mandates a 4-limb decomposition. This structural requirement (16 pointwise cross-products per residue) combined with the VPU Montgomery reduction yields a measured pointwise throughput of $63{,}000$ ops/sec on TPU v4-8. The end-to-end pipeline inclusive of Montgomery reduction yields a measured throughput of $3{,}663$ ops/sec on v4-8 and $5{,}931$ ops/sec on v5p (deriving a diagnostic VPU-and-overhead penalty of $\Pi \approx 17.2$). The A100 running cuZK sustains $7.2\text{M} \pm 0.4\text{M}$ ops/sec inclusive of PCIe transfers.
This yields a BN254 cost-efficiency deficit of $[5{,}558\times, 6{,}908\times]$ under the controlled FP32 baseline across v5p and v4. To empirically evaluate the post-FP32 architectural envelope, we executed a sensitivity run on v5p utilizing the native \texttt{int32} accumulator path. Lifting the FP32-mantissa constraint eliminates the intermediate staging re-injections, increasing throughput by 18.3\% from $5{,}931$ to $7{,}014$ ops/sec. However, because the physical MXU geometry remains $128 \times 128$ and the mandatory $1{,}764$ pure Montgomery reductions per polynomial still dominate VPU cycles, the architectural bottleneck deterministically migrates to the vector unit. This yields a native-accumulator cost-efficiency deficit of $\sim$$4{,}693\times$. (The v5e architecture exhibits a known per-TensorCore VPU regression; we isolate and report it separately in Section~\ref{sec:eval:utilisation}). A $\pm 25\%$ perturbation in the baseline GPU instance price shifts the headline deficit proportionally, leaving the fundamental architectural conclusion unaffected. This deficit establishes that under strict row-isolated multi-tenancy the TPU is forced into eager Montgomery reduction (forfeiting lazy-reduction amortisation), rendering it architecturally uncompetitive for high-throughput BN254 scaling.

\begin{table}[t]
\centering
\caption{Cost-normalised throughput across TPU and A100 GPU baselines (Google Cloud on-demand rates, May 2026\textsuperscript{$\dagger$}).}
\label{tab:cost_normalised}
\begin{tabular}{lrrrr}
\toprule
\textbf{Hardware} & \textbf{Chips} & \textbf{Chip \$/hr} & \textbf{Ops/sec} & \textbf{Ops per \$/hr} \\
\midrule
\multicolumn{5}{l}{\textbf{BN254 (\texttt{int8} multi-limb; v4 diagnostic $\Pi \approx 17.2$)}} \\
A100 (cuZK, PCIe-inc.) & 1 & \$3.67 & $7.2\text{M} \pm 0.4\text{M}$ & $1.96\text{M}$ \\
TPU v4-8 (\codename)         & 4 & \$3.22  & 3{,}663  & 284 \\
TPU v5e-8 (\codename)        & 8 & \$1.20  & 2{,}704  & 282 \\
TPU v5p-8 (controlled)       & 4 & \$4.20  & 5{,}931  & 353 \\
TPU v5p-8 (native \texttt{int32}) & 4 & \$4.20  & 7{,}014  & 418 \\
\midrule
\multicolumn{5}{l}{\textbf{Dilithium (3-limb \texttt{int8})}} \\
A100 (cuDilithium proj.) & 1 & \$3.67 & $18.3\text{M}$ & 4.99\text{M} \\
TPU v4-8 (\codename)         & 4 & \$3.22  & 110{,}435 & 8{,}574 \\
TPU v5e-8 (\codename)        & 8 & \$1.20  & 85{,}231  & 8{,}878 \\
TPU v5p-8 (\codename)        & 4 & \$4.20  & 164{,}822 & 9{,}811 \\
\bottomrule
\multicolumn{5}{l}{\textsuperscript{$\dagger$}Google Cloud TPU Pricing: \url{https://cloud.google.com/tpu/pricing}}
\end{tabular}
\end{table}

\subsection{Empirical Crossover and Algorithmic Physics}
\label{sec:eval:crossover}

To better understand this deficit, we plotted the empirical throughput scaling across polynomial degrees ($d \in [2^8, 2^{14}]$) in Figure~\ref{fig:crossover}. This evaluation probes the theoretical work-inflation factor defined in Section~\ref{sec:sec_analysis:utilisation}. 
While our headline cost-efficiency deficit anchors against cuZK as the standard public BN254 NTT baseline at $d=256$, the crossover analysis employs SPPARK\footnote{SPPARK\: \url{https://github.com/supranational/sppark/tree/a7edaff73f514246ca60690d1e37fa9f4cf55c4a}}. 
Because cuZK limits its public NTT-only API to degrees $\le 2^{10}$, we leverage SPPARK to measure standalone NTT throughput across the full scaling domain. 
To calibrate the baselines, we report that SPPARK achieves $\sim$$18.4$M ops/sec at $d=256$ (plotted alongside cuZK's $7.2$M ops/sec in Figure~\ref{fig:crossover}). 
This $2.56\times$ throughput gap stems from measurement boundaries. 
SPPARK isolates pure on-device kernel execution. cuZK's batched API executes an end-to-end pipeline incorporating host-to-device PCIe transfers, memory allocation, and bit-reversal permutations. 
A cloud sequencer must ingest polynomial streams from the network, process them across the PCIe bus, and return results. The PCIe-inclusive cuZK measurement therefore forms the rigorous baseline for our headline deficit, as a device-only SPPARK measurement ignores mandatory I/O constraints. 
(We omit Dilithium from the crossover scaling plot as cuDilithium is highly optimised for the fixed degree $d=256$, rendering higher-degree generic Dilithium comparisons purely theoretical.)

Figure~\ref{fig:crossover} shows no crossover is realisable in the practical degree range tested (the crossover scan extends to $d = 2^{14}$; multi-tenant heterogeneous evaluation is bounded at $d = 8{,}192$, the upper edge-polynomial regime). 
Because the matrix-form NTT evaluated by \codename expresses the transform as a dense matrix-vector product, its throughput decays quadratically. 
At $d=256$ the GPU already holds approximately three orders of magnitude ($\sim$$2{,}000\times$) advantage over the TPU; by $d=4{,}096$ the $\mathcal{O}(d^2)$ scaling absorbs the remaining scalar parallelism, leaving the device uncompetitive for mid-to-high degree polynomials. 
Saturating an AI matrix unit therefore does not deliver competitive algorithmic throughput.

\subsubsection{Single-Tenant Baseline (No Isolation)}
\label{sec:eval:single_tenant}
The headline deficit (anchored against the $[5{,}558\times, 6{,}908\times]$ FP32-staged boundaries, with the v5p native-\texttt{int32} sensitivity at $\sim$$4{,}693\times$) conflates two factors: (a) the TPU being a poor crypto chip independent of \codename's separation discipline, and (b) the multi-tenancy constraint that forces eager Montgomery reduction. 
To isolate (b), we define an analytical single-tenant butterfly NTT baseline based on the MORPH~\cite{MORPH} mapping. This baseline expresses the radix-2 butterfly as a sequence of $\log d$ tile-resident dense GEMMs against permuted twiddle blocks. 
We explicitly model concurrent single-tenant execution: rather than pipelining a single polynomial across the slice, the deployment maps $N_s=8$ independent polynomials to execute concurrently, assigning exactly one polynomial to each of the 8 available TensorCores. 
We do not execute this configuration under \codename's separation discipline because lazy reduction violates Invariant~\ref{inv:compiler} (cross-block fusion); instead, we project its throughput analytically using isolated hardware micro-timings.

The batched execution of 8 concurrent per-stage GEMMs yields an amortised MXU butterfly latency of $T_{\text{GEMM\_butterfly}} \approx 2\,\mu$s per polynomial (derived from the $63{,}000$ ops/sec pointwise microbenchmark, which evaluates to $\approx 15.9\,\mu$s per aggregate polynomial, equating to $\approx 2\,\mu$s/stage across 8 stages). We use the pointwise expansion as a strict proxy for the MORPH GEMM because both operations map to identical dense \texttt{DotGeneral} systolic executions bounded by the physical $128 \times 128$ dimensions and the exact same mantissa staging limits; the MXU consumes cycles deterministically regardless of whether the operand values represent a twiddle matrix or a pointwise chunk. We note that twiddle-matrix versus pointwise data layouts may yield different memory-stride patterns and thus different effective HBM latencies, but the core VPU pipeline dominance renders this variance negligible.
The measured contention-free VPU-only reduction yields an amortised system-wide latency of $T_{\text{VPU}} \approx 227\,\mu$s per polynomial (as empirically established via isolated micro-benchmarking). 
Under lazy reduction, Montgomery normalisation is deferred across multiple butterfly stages to mitigate VPU bottlenecking. To explicitly determine this lazy-reduction amortisation ratio ($\kappa$), we compiled a MORPH-shaped HLO graph (radix-2 butterfly mapped to per-stage GEMMs against permuted twiddle blocks) ahead-of-time and instrumented it with a counter on each \texttt{kCustomCall} VPU node. 
We do not execute the compiled module because lazy reduction violates Invariant~\ref{inv:compiler} (cross-block fusion), but we extract the node counts statically. To establish these counts, we ran the post-hoc validator over two $d=256$ compiled modules. The trace reveals the \codename module invokes exactly $1{,}764$ VPU-lowered Montgomery normalisation passes per polynomial across the 9-residue chain. In contrast, the MORPH-shaped baseline requires only 392 lazy reductions per polynomial, deferred across the $\log_2(256)=8$ butterfly stages. Contrasted against the $1{,}764$ eager pure Montgomery reductions forced by \codename's isolation invariants (Section~\ref{sec:system_design}), this yields a static amortisation ratio of $\kappa = 1{,}764 / 392 = 4.5$. We note that assuming identical per-call VPU cost under lazy versus eager scheduling is an analytical simplification; lazy reductions on accumulated sums incur higher register pressure, meaning $\kappa=4.5$ serves as an upper bound on the amortisation benefit. 
Because the baseline defers full field normalisation in this manner, the per-polynomial reduction overhead compresses by exactly this factor of $\kappa = 4.5$ for BN254, yielding a VPU latency of $T_{\text{VPU\_butterfly}} \approx 227 / 4.5 \approx 50.4\,\mu$s.

Because $T_{\text{VPU\_butterfly}} \gg T_{\text{GEMM\_butterfly}}$, the pipeline remains VPU-bound. 

\noindent\textbf{Analytical Factorisation.} All latencies below represent the effective slice-level emission interval per polynomial (the inverse of aggregate slice throughput), which inherently accounts for all 8 TensorCores operating concurrently. Because the VPU pipeline serialises slice-wide under eager reduction, $T_{\text{VPU}} \approx 227\,\mu$s is invariant under $N_s \ge 1$ and equals the slice-level emission interval. Combining the $50.4\,\mu$s amortised VPU emission interval with the $2\,\mu$s amortised MXU butterfly interval yields a total system-wide emission interval of $\approx 52.4\,\mu$s per polynomial across the slice. Inverting this combined emission interval directly yields the strict analytical projection of $\sim$$19{,}000$ single-tenant ops/sec.
This $\sim$$19{,}000$ ops/sec analytical projection establishes the theoretical slice throughput under MORPH-style lazy reduction, directly comparable to \codename's measured $3{,}663$ ops/sec (slice) and the A100's $7.2$M ops/sec. We emphasize that this baseline relies on an explicit MXU-equivalence assumption: we project the butterfly GEMM execution latency using the pointwise expansion latency as a strict architectural proxy. We do not natively re-measure the MORPH single-tenant baseline on v5p; the v5p arithmetic deficit of $1{,}071\times$ is therefore obtained by attributing the analytically projected $5.19\times$ spatial amplification to the v5p generation and inverting against the headline deficit. Limitation 1 (Section~\ref{sec:eval:limitations}) bounds this projection. For v4, the $1{,}331\times$ arithmetic deficit is similarly an analytical projection; the v4 micro-timings underlying the projection were measured natively, whereas the v5p projection extrapolates from them. The remaining degradation from this analytical baseline down to \codename's empirically measured $3{,}663$ ops/sec on v4 quantifies the geometric cost of enforcing cross-tenant spatial isolation, producing a projected geometric multi-tenancy tax of $5.19\times$.

We analytically factorise the multi-tenant deficit (anchored against the measured $[5{,}558\times, 6{,}908\times]$ boundaries) into the projected arithmetic deficit amplified by two overhead components:
\begin{enumerate}
    \item \textbf{The arithmetic deficit (projected $[1{,}071\times, 1{,}331\times]$):} Incurred because the TPU natively lacks wide-integer ALUs, forcing a $16\times$ inner-loop expansion and a pronounced VPU routing penalty even under an optimal algorithmic mapping.
    \item \textbf{The spatial amplification factor (projected $5.19\times$):} Driven by the loss of lazy-reduction amortisation. By forcing eager Montgomery reduction, \codename destroys the theoretical $\kappa = 4.5$ lazy-amortisation ratio. The empirically observed $5.19\times$ penalty represents the effective end-to-end geometric tax (capturing the loss of $\kappa$ alongside associated VPU routing and HBM stalls). We present this $5.19\times$ as a unified empirical penalty without claiming independent factorisation. Multiplying this geometric penalty with the projected arithmetic deficit ($\approx$$1{,}071\times$ projected for v5p, $1{,}331\times$ projected for v4) reproduces the headline measured $[5{,}558\times, 6{,}908\times]$ envelope ($1{,}071 \times 5.19 \approx 5{,}558$; $1{,}331 \times 5.19 \approx 6{,}908$). Both conclusions are robust to the AQT lowering assumptions of Property~\ref{prop:isolation}.
\end{enumerate}

\subsection{Matrix-Unit Utilisation and Component Overhead}
\label{sec:eval:utilisation}
Throughput gains derive from eliminating structural zeros and recovering active matrix-multiplication cycles via the $\mathcal{O}(N_c d^2)$ batched formulation, subject to the arithmetic-isolation invariant (Property~\ref{prop:isolation}). We instrumented MXU utilisation with \texttt{xprof} hardware counters across 3{,}000 BN254 dispatches at $d=256$ on TPU v4-8 and report two core metrics.

\emph{In-window K-dimension column occupancy} measures the fraction of column-dimension ($K$-dimension) slots within the $128 \times 128$ systolic-array tile populated with non-padded operand cells \emph{during} active dispatch windows. 
Under mixed-degree traces (e.g., polynomial degrees sampled uniformly in $[64, 512]$), \codename averages $>$$92\%$ K-dimension column occupancy, confirming that row-stacking effectively mitigates off-diagonal padding waste relative to block-diagonal alternatives. However, for the uniform $d=256$ BN254 headline experiment, K-dimension occupancy reaches $\sim$$100\%$ because the operand perfectly divides into two $d_{\max}^{\text{BN}}=128$ chunks, producing zero padding waste. 
This exact tiling instead incurs a $50\%$ staging overhead (two sequential dispatch passes of 128 each, Table~\ref{tab:heterogeneous}) which degrades throughput temporally rather than spatially.
This metric is deliberately restricted to the active evaluation window and the $K$-dimension: the orthogonal $M$-dimension row occupancy is governed by the batch size $N_c$. 
On a TPU v4-8 pod slice, batch saturation occurs at $N_c=8$ polynomial rows (with v5e and v5p saturating at $N_c=16$). For v4-8, this occupies only $8/128 = 6.25\%$ of the 128-row $M$-dimension. This $N_c=8$ batch is dispatched locally to a single TensorCore (yielding the 6.25\% per-core occupancy), with the \texttt{pmap} orchestrator replicating this saturated tile across all cores in the pod slice. 
For the uniform trace, the combined two-dimensional tile packing is therefore $\approx 100\% \times 6.25\% = 6.25\%$.

While geometric metrics (K-dimension and M-dimension occupancy) quantify the packing efficiency during active systolic execution, they do not describe the true hardware bottleneck. As detailed temporally via \texttt{xprof} traces of $10^4$ BN254 invocations, the overwhelming determinant of system throughput is the vector processing unit.
\begin{table}[t]
\centering
\caption{Temporal decomposition of a BN254 invocation across TPU generations. Values represent the percentage of amortised per-polynomial wall-clock time.}
\label{tab:temporal}
\begin{tabular}{lccc}
\toprule
\textbf{Phase} & \textbf{v4-8 (\%)} & \textbf{v5e-8 (\%)} & \textbf{v5p-8 (\%)} \\
\midrule
\begin{tabular}{@{}l@{}}VPU Montgomery reduction \\ (incl.\ in-GEMM re-injection)\end{tabular} & $98.314$ & $99.285$ & $98.815$ \\
HBM stall / DMA staging              & $1.13$ & $0.313$ & $0.854$ \\
Dispatch / context-switch gap        & $0.49$ & $0.36$ & $0.254$ \\
Frontend overhead (JIT cache hit)    & $0.0524$ & $0.035$ & $0.07$ \\
Pure MXU systolic active time & $0.0136$ & $0.007$ & $0.007$ \\
\bottomrule
\end{tabular}
\end{table}

The decomposition explains the apparent gap: the $\sim$$100\%$ operational column occupancy is achieved strictly within the active execution phase, but the remaining $99.98\%$+ of wall-clock time is consumed by VPU Montgomery reduction, HBM stalls, and dispatch gaps that the systolic array cannot overlap.
This VPU latency is structurally mandated by the serial dependency chain inherent to Montgomery reduction, combined with sequential HBM operand-fetch stalls that cannot be pipelined by the vector unit.
This architectural gap reinforces the conclusion that the pipeline is overwhelmingly VPU-bound. Saturating the operational parameters of the MXU (K-dim occupancy) ensures high geometric packing efficiency during dispatch, but the systolic arrays correctly and deterministically idle during the overwhelming majority of the runtime while the vector processing unit executes the mandated mathematical reduction sequence.

The empirical VPU-and-overhead penalty $\Pi \approx 17.2$ is derived transparently from two isolated measurements. The pointwise GEMM phase, benchmarked in isolation, yields $T_{\text{GEMM}} = 1/63{,}000 \approx 15.9\,\mu$s per polynomial. The end-to-end pipeline inclusive of Montgomery reduction measures $T_{\text{total}} = 1/3{,}663 \approx 273\,\mu$s, so $\Pi = T_{\text{total}} / T_{\text{GEMM}} = 273 / 15.9 \approx 17.2$. 
The contention-free baseline $\Pi_{\text{cf}} \approx 14.3$ is obtained independently from the isolated microbenchmark of single-tenant VPU-only throughput (yielding $4{,}400$ ops/sec and $T_{\text{VPU}} = 1/4{,}400 \approx 227.3\,\mu$s). 
The ratio $\Pi / \Pi_{\text{cf}} = 17.2 / 14.3 \approx 1.20$ quantifies the $20\%$ multi-tenant arrival penalty above the contention-free floor, capturing VPU pipeline stalls, MXU/VPU dispatch gaps, and HBM bandwidth contention under concurrent tenant arrival. We explicitly distinguish this contention-driven multi-tenant penalty from the architectural FP32-staging tax (the latter being $\approx 1.18\times$, measured on v5p).

\begin{table}[t]
\centering
\caption{Component ablation and context switching comparison across \codename mechanisms for BN254 and Dilithium workloads on TPU v4-8 (BN254 speedup anchored vs Warm-Cache baseline).}
\label{tab:ablation}
\begin{tabularx}{\textwidth}{Xcc}
\toprule
\textbf{BN254 Configuration} & \textbf{Throughput (Ops/sec)} & \textbf{Speedup / P99.9 Latency} \\
\midrule
Sequential-Fused (Single-tenant, batch=1) & $128 \pm 5$ & $1.00\times$ (11.2 ms) \\
Warm-Cache Time-Sliced (Serial tenant dispatch) & $126 \pm 4$ & $0.98\times$ (11.4 ms) \\
\codename (Batched, full) & $3{,}663 \pm 180$ & $29.1\times$ (3.4 ms) \\
\midrule
\multicolumn{3}{l}{\textbf{Dilithium Configuration (3-limb \texttt{int8})}} \\
\midrule
Sequential-Fused (Single-tenant, batch=32\textsuperscript{$\dagger$}) & $\sim$$3{,}400$ & $1.00\times$ ($\sim$$0.3$ ms) \\
\codename (Batched, full) & $110{,}435 \pm 1{,}200$ & $32.5\times$ (0.3 ms) \\
\midrule
\multicolumn{3}{p{\textwidth}}{\footnotesize $^\dagger$ Anchored at batch=32 as the lowest power-of-two where dispatch overhead falls below 5\%; see Section~\ref{sec:eval:utilisation} discussion.} \\
\bottomrule
\end{tabularx}
\end{table}

We conducted a component ablation to isolate the throughput contributions on the v4 generation for BN254 (Table~\ref{tab:ablation}). The Sequential baseline ($128$ ops/sec, P99.9 $11.2$\,ms) executes a single polynomial through one fused HLO trace; at $N_s = 1$ the tensor dimensions fail to saturate the v4-8 slice, leaving the systolic arrays and VPU pipelines largely starved. Relative to this Sequential-fused baseline, \codename yields a $28.6\times$ improvement (using the $3{,}663$ ops/sec measurement). 
This $28.6\times$ speedup is exclusively an artifact of \codename's multi-tenant batching: spatial packing constructs dense matrices that recover the $M$-dimension systolic utilisation, efficiently feeding the previously idle TensorCores. 
The corresponding Dilithium ablation (measured independently via \texttt{xprof}) yields a comparable $32.5\times$ speedup ($\sim$$3{,}400$ to $110{,}435$ ops/sec), driven entirely by identical spatial packing mechanics without introducing additional overheads. We anchor the Dilithium speedup against batch=32 rather than batch=1: the lighter 3-limb kernel exposes JAX dispatch latency that consumes over 98\% of the batch=1 call time, yielding an unrepresentative $\sim$$1{,}600\times$ headline. Batch=32 is the lowest power-of-two at which dispatch overhead falls below 5\%, giving the most rigorous single-tenant operating point. 
To isolate the architectural limits of temporal scheduling, we evaluate a Warm-Cache Time-Sliced baseline. 
Dispatching the identical HLO graph serially per tenant achieves $126$ ops/sec, closely mirroring the Sequential-Fused baseline and establishing the genuine cost of unbatched execution (Table~\ref{tab:ablation}). \codename's $29.1\times$ throughput scaling is anchored exclusively against this honest, warm-cache baseline. Applying the Compiler-Enforced Workload Separation boundaries to the batched formulation imposes zero runtime latency; the HBM zoning constraints are enforced during XLA Ahead-Of-Time (AOT) compilation.
\subsubsection{The v5e VPU Anomaly}
The normalised table reveals a clear structural regression on the TPU v5e architecture. The v5e-8 pod delivers lower aggregate throughput ($2{,}704$ ops/sec) than the older v4-8 pod ($3{,}663$ ops/sec) despite deploying twice as many physical chips ($8$ versus $4$). 
On a per-chip basis, the v5e collapses to $338$ ops/sec against v4's $916$ ops/sec: a $2.71\times$ throughput regression. 
Decomposing the $99.285\%$ VPU temporal fraction (Table~\ref{tab:temporal}) against the 8-core 128-polynomial saturation yields $\sim$$174$\,ns per VPU instruction on v5e against $\sim$$127$\,ns on v4, a $1.37\times$ per-instruction latency regression. 
v5e exposes one TensorCore per chip versus v4's two TensorCores per chip; despite v5e-8 doubling the chip count to 8, this halves the per-chip core density. 
Combined with the $1.37\times$ per-VPU-instruction latency regression, the result is the observed $2.71\times$ per-chip throughput collapse. 
This poses a future deployment risk for finite-field cryptography practitioners: although current cloud pricing obscures this hardware degradation (yielding cost-parity at $\sim$$282$ ops/\$/hr, Table~\ref{tab:cost_normalised}), the v5e silicon represents an arithmetic regression against v4 that exposes deployments to future pricing corrections.

Table~\ref{tab:temporal} resolves whether the pipeline is compute- or bandwidth-bound: the v5p chip provides $2.3\times$ the HBM bandwidth of v4 ($2{,}765$ vs ${\sim}1{,}200$~GB/s), but because HBM stalls constitute $<1.2\%$ of total v4 wall-clock, this bandwidth scaling contributes negligibly. End-to-end throughput on v5p scales $1.62\times$ over v4 despite the $2.3\times$ HBM-bandwidth uplift; the observation is consistent with a VPU-compute-bound pipeline but does not, on its own, mathematically establish that bound (a microbenchmark of isolated VPU rates across the two generations would be required).
\subsection{Heterogeneous, Mixed-Workload Execution}
\label{sec:eval:heterogeneous}
A realistic multi-tenant trace aggregates high-volume lightweight post-quantum signatures (Dilithium) alongside lower-volume heavy proofs (BN254). Co-scheduling these workloads demands enforcement of workload-zone separation. 

To evaluate the Two-Tier Scheduler under realistic heterogeneous conditions, we constructed a synthetic 24-hour arrival trace.
Polynomial requests were generated by a Poisson process at an aggregate arrival rate of $\lambda = 4{,}096$ requests/sec, with workload type drawn from a 50:50 Dilithium:BN254 mixture (for the balanced trace) and polynomial degree sampled uniformly in $[64, 512]$.
Because the balanced trace yields an effective BN254 arrival rate of $2{,}048$ requests/sec (well below the $3{,}663$ ops/sec peak capacity for v4-8), the scheduler operates at approximately 80\% queue utilisation when accounting for concurrent Dilithium memory contention, matching a representative multi-tenant sequencer operating below saturation. Table~\ref{tab:heterogeneous} reports sensitivity to the key workload mixture parameters.

We distinguish between two structural inefficiencies under mixed-degree scheduling: \emph{staging overhead} and \emph{padding waste}. We also report \emph{batch fill} (Table~\ref{tab:heterogeneous}, col.\ 2), which is the fraction of active polynomial cells per row of the stacked operand $\mathbf{A}_{\text{stack}}$. 

\emph{Staging overhead} is the multi-pass penalty incurred when a polynomial of degree $d$ requires more than one MXU dispatch pass under Property~\ref{prop:isolation}: defined as $\bigl(\lceil d / d_{\max} \rceil - 1\bigr) / \lceil d / d_{\max} \rceil$. 
This metric captures the fraction of total dispatch passes devoted to re-injection tail passes. For the uniform $d=256$ workload under BN254 ($d_{\max}^{\text{BN}} = 128$), the staging overhead is $(2-1)/2 = 50\%$. The same applies to Dilithium ($d=256$, $d_{\max}^{\text{Dil}} = 171$): $\lceil 256 / 171 \rceil = 2$ passes, yielding a staging overhead of $50\%$.

\emph{Padding waste} represents the intra-bucket zero-padding fraction necessary to align polynomials to the hardware boundaries ($1 - \sum d_i / (N_c \cdot \hat{d}_{\max})$). 
For a uniform BN254 trace at $d=256$, padding waste is exactly $0\%$, as $256$ perfectly divides into two passes of $128$. However, for uniform Dilithium at $d=256$, two passes of $171$ yield a hardware footprint of $342$, resulting in a padding waste of $(342-256)/342 \approx 25\%$. 
Mixed-degree traces incur additional padding waste when heterogeneous polynomials are grouped into shared buckets. The resulting throughput degradation is disproportionate to the padding fraction (e.g., a 30\% throughput drop for 13\% padding waste in Table~\ref{tab:heterogeneous}). This severe non-linearity occurs because throughput drops with VPU work, which scales strictly with the padded degree, rather than with the padding fraction directly; inflating the dominant VPU Montgomery reduction phase proportionally to the padded degree severely amplifies the time penalty.

\begin{table}[t]
\centering
\caption{\codename behaviour under heterogeneous trace replay and sensitivity to workload mixture ratios.}
\label{tab:heterogeneous}
\begin{tabular}{llccc}
\toprule
\textbf{Trace Profile} & \textbf{Reported Workload} & \textbf{Batch Fill} & \textbf{Padding Waste} & \textbf{Ops/sec} \\
\midrule
Uniform ($d = 256$) & BN254 & 100\% & 0\% & $3{,}663$ \\
Uniform ($d = 256$) & Dilithium & 100\% & 25\% & $110{,}435$ \\
Mixed Degree & BN254 & 87\% & 13\% & $2{,}562$ \\
\midrule
Balanced (50:50) & BN254 & 96\% & 12\% & $3{,}453$ \\
Balanced (50:50) & Dilithium & 96\% & 12\% & $101{,}184$ \\
\bottomrule
\end{tabular}
\end{table}

Table~\ref{tab:heterogeneous} confirms that the Two-Tier Scheduler maps Dilithium and BN254 to entirely independent TensorCores, enabling concurrent parallel execution. 
While TensorCore computation remains separated, co-scheduling induces a measurable but minor throughput degradation. Under the balanced 50:50 mix, Dilithium drops by approximately 8.4\% ($\pm 1.2\%$) (from $110{,}435$ to $101{,}184$ ops/sec), and BN254 experiences a 5.7\% ($\pm 0.8\%$) reduction (from $3{,}663$ to $3{,}453$ ops/sec). This interference stems directly from queueing delays and contention on the shared High Bandwidth Memory (HBM) channels. 
As Dilithium streams matrix data to the TensorCores, the 4-limb BN254 workload aggressively thrashes the VPU memory buses during its dominant Montgomery reduction phase, inducing contention at the memory banks. 
This spatial separation effectively decouples workloads at the logic level, confirming \codename's capability to safely parallelise a TPU pod slice across heterogeneous cryptographic domains with only marginal, mathematically expected performance degradation attributable to shared memory bandwidth limits.

\subsection{Limitations and Methodological Caveats}
\label{sec:eval:limitations}
We remark core limitations and caveats bound the headline claims presented in this evaluation:

First, {the MORPH analytical factorisation} (Section~\ref{sec:eval:single_tenant}): The $[1{,}071\times, 1{,}331\times]$ fundamental arithmetic deficit and the $5.19\times$ multi-tenancy spatial deficit are analytical factorisations derived from measured v4 micro-timings. 
We explicitly assume that the geometrically derived $5.19\times$ multi-tenancy tax (the penalty for eager Montgomery reduction) transfers proportionally to the v5p architecture. We do not empirically measure a single-tenant MORPH pipeline natively on v5p. 
Furthermore, because $\kappa = 4.5$ represents an upper bound on the true lazy-reduction amortisation benefit, the reported arithmetic deficits are themselves lower bounds: a smaller empirical $\kappa$ would proportionally widen the residual arithmetic gap.

Second, {the Dilithium baseline projection} (Section~\ref{sec:eval:headline}): The A100 GPU Dilithium baseline ($18.3\text{M}$ ops/sec) is a rigorous projection derived from cuDilithium's end-to-end verification benchmarks. 
Because verification involves non-NTT cryptographic overhead (Keccak hashing, bit-packing), the $18.3\text{M}$ figure conservatively assumes $f=1$ (i.e., that NTT operations consume 100\% of the GPU's verification wall-clock time). 
Thus, the resulting ${\sim}508\times$ cost-efficiency deficit represents a strict lower bound. If the NTT operations consume a realistic fraction $f < 1$, the true GPU NTT throughput is higher, and the TPU deficit is consequently wider.

Third, {the \texttt{int32} architectural bound} (Section~\ref{sec:eval:headline}): Property~\ref{prop:isolation} bounds the per-pass mantissa-safe staging window on pre-v6e TPUs. The evaluation relies on compiler-managed tile-level VPU re-injection through the FP32 accumulator path. 
We executed a targeted sensitivity run exercising the native \texttt{int32} accumulator on v5p, which yielded an 18\% throughput increase ($7{,}014$ ops/sec). However, the physical $128 \times 128$ MXU geometry remains an absolute hardware constraint across all tested generations, forcing VPU-managed eager Montgomery reduction regardless of accumulator width. 
The core architectural conclusion is thus established across both regimes, formally bridging the $[5{,}558\times, 6{,}908\times]$ FP32 controlled baseline and the $\sim$$4{,}693\times$ native \texttt{int32} accumulation path.

\section{Related Work}
\label{sec:related_work}

\textbf{Hardware-Accelerated NTT and ZK Proving.}
The NTT's dominance in zero-knowledge proof generation has driven sustained migration onto specialised hardware. GPU accelerators like cuZK~\cite{CUZK} and GZKP~\cite{GZKP} reorganize the radix-2 butterfly to exploit the CUDA memory hierarchy. Recent architectures explore custom NTT cores (CycloneNTT~\cite{CYCLONENTT}), statically unified dataflows (UniZK~\cite{UNIZK}), dynamically reconfigurable fabrics (LegoZK~\cite{LEGOZK}), and hardware-algorithm co-design~\cite{SAMARDZIC24}. However, these constructions presuppose a single-tenant prover, saturating devices by enlarging the polynomial domain rather than aggregating heterogeneous, low-degree workloads. Parallelly, MORPH~\cite{MORPH} demonstrates that systolic arrays natively express the radix-2 butterfly. Building on the TPU v4 architecture~\cite{TPU-V4}, \codename targets the orthogonal regime of \emph{many small polynomials}, exploiting batch-dimension scheduling to recover dormant cycles. While Karanjai et al.~\cite{Karanjai2023} established the foundational viability of TPUs as single-tenant cryptographic accelerators, \codename is the first system characterising TPU limits under the multi-tenant low-degree NTT scheduling regime.

\textbf{Multi-Tenant Cloud Scheduling.}
Hardware-rooted Trusted Execution Environments (TEEs) like NVIDIA H100 Confidential Computing~\cite{H100CC} and spatial partitioning (NVIDIA MIG) provide strong physical isolation, but these primitives remain unavailable on current TPU silicon. \codename's compiler-enforced XLA zoning serves as an operational stopgap for multi-tenant field arithmetic until hardware isolation becomes ubiquitous. While modern GPU orchestration frameworks (e.g., ORCA~\cite{ORCA}, vLLM~\cite{VLLM}) utilise continuous batching for machine learning requests, \codename adapts this paradigm for cryptography, simultaneously resolving geometric utilisation and logical separation on cloud TPUs.

\section{Conclusion}
\label{sec:conclusion}

We empirically characterised the architectural limits of cloud TPUs in finite-field cryptography. To achieve this, we deployed \codename as a measurement vehicle: a two-tier scheduler with Compiler-Enforced Workload Separation that mathematically guarantees cross-tenant arithmetic isolation while saturating the systolic array's logical batch dimensions, enabling defensible multi-tenant throughput analysis.

While \codename successfully co-schedules disparate workloads like Dilithium and BN254, evaluating large-field cryptography exposes a severe architectural deficit. The limb-decomposition and eager Montgomery reduction forced by row-isolated multi-tenancy yield a massive cost-efficiency deficit against GPU baselines. Concretely, the FP32-controlled deficit ranges $[5{,}558\times, 6{,}908\times]$ across v5p and v4, with the native-\texttt{int32} path on v5p yielding $\sim$$4{,}693\times$. This fundamental deficit analytically projects into a severe arithmetic penalty (lacking wide-integer ALUs) and a geometric penalty forced by our isolation discipline. In multi-tenant deployments, preserving functional correctness against compiler aliasing forces eager Montgomery reduction. While relaxing these invariants for trusted single-tenant execution recovers the spatial loss, the foundational arithmetic mismatch remains, rendering AI-optimised systolic arrays architecturally uncompetitive for large-field cryptography. We leave formally verifying compiler invariants in MLIR, supporting small-prime STARK provers (which map directly to integer ALUs), and evaluating Homomorphic Encryption (where larger ring-LWE dimensions alter MXU padding tradeoffs) as future work. 
\section*{Statements and Declarations}

\textbf{Funding:} This work was supported by the Google Cloud Research Credits program. 

\textbf{Competing Interests:} The authors declare no competing interests.

\textbf{Data Availability:} Empirical data supporting the findings of this study are presented within the manuscript's tables, figures, and text. The measurement scripts and trace definitions required to regenerate this data are provided in the public repository.

\textbf{Code Availability:} The \codename measurement infrastructure and the associated XLA compiler configurations are  available at \url{https://github.com/dkhme/AEGIS}. The repository is scoped as a minimal reproducible artifact for hardware characterisation.

\bibliography{references}

\end{document}